\documentclass[11pt]{article}

 \usepackage{amsmath,amsthm,amssymb}
 \usepackage{makeidx,epsfig,lscape}
 \usepackage{color,colortbl}
 \usepackage{fancyhdr}
 \usepackage{xcolor,pict2e}

\usepackage{lineno}
\usepackage{physics}
\usepackage{stackengine}
\usepackage{scalerel}
\usepackage{esvect}
\usepackage{enumitem}
 \renewcommand{\headrulewidth}{0pt}
 \renewcommand{\footrulewidth}{0.5pt}

 \definecolor{myaqua}{rgb}{0.0,0.5,0.55}
 \definecolor{lightaqua}{rgb}{0.75,0.95,0.95}
  \definecolor{maroon}{rgb}{0.824,0.137,0.169}

 \usepackage[colorlinks = true,
            linkcolor = myaqua,
            urlcolor  = blue,
            citecolor = myaqua]{hyperref}

\usepackage{caption}
\usepackage{floatrow}

\newcommand{\xhat}{\boldsymbol{\hat{\textbf{\textit{x}}}}}
\newcommand{\yhat}{\boldsymbol{\hat{\textbf{\textit{y}}}}}
\newcommand{\zhat}{\boldsymbol{\hat{\textbf{\textit{z}}}}}

\def\lin#1#2{\textcolor[rgb]{0.6,0.6,0.6}{\vspace*{#1mm} \hrule
   height 3 pt \vspace*{#2mm}}}
\def\bt{\begin{tabular}}
\def\et{\end{tabular}}
\def\and{\mbox{ and }}

\def\P{\mbox{\bf P}}

\def\1{{\bf 1}}

 \def\sectionn#1{\refstepcounter{section}{\color{maroon}

 \vskip 6mm

 \noindent\Large\bf\thesection. #1}

 \vskip 3mm}
 \def\subsectionn#1{\refstepcounter{subsection}{\color{myaqua}

 \vskip 5mm

 \noindent\large\bf\thesubsection. #1}

 \vskip 2mm}

\begin{document}

 \fancyhead[L]{\hspace*{-13mm}
 \bt{l}{\bf Open Journal of *****, 2019, *,**}\\
 Published Online **** 2019 in SciRes.
 \href{http://www.scirp.org/journal/*****}{\color{blue}{\underline{\smash{http://www.scirp.org/journal/****}}}} \\
 \href{http://dx.doi.org/10.4236/****.2019.*****}{\color{blue}{\underline{\smash{http://dx.doi.org/10.4236/****.2019.*****}}}} \\
 \et}

 $\mbox{ }$
 \vskip 12mm

{
{\noindent{\huge\bf Probing entanglement in Compton interactions}}
%
\\[6mm]
{\large\bf Peter Caradonna\textcolor{myaqua}{$^{1}$$^{*}$}, David Reutens\textcolor{myaqua}{$^{1,2}$}, Tadayuki Takahashi\textcolor{myaqua}{$^{3}$}, Shin'ichiro Takeda\textcolor{myaqua}{$^{3}$}, Viktor Vegh\textcolor{myaqua}{$^{1,2}$}}}
\\
{$^1$Centre for Advanced Imaging, The University of Queensland, Brisbane 4072, Australia\\
$^2$ARC Training Centre for Innovation in Biomedical Imaging Technology, Australia\\
$^3$Kavli Institute for the Physics and Mathematics of the Universe (WPI), Institutes for Advanced Study (UTIAS), The University of Tokyo,
5-1-5 Kashiwa-no-Ha, Kashiwa, Chiba, 277-8583, Japan}\\

\noindent{\textcolor{myaqua}{$^{*}$} Present address: Kavli Institute for the Physics and Mathematics of the Universe (WPI), Institutes for Advanced Study (UTIAS), The University of Tokyo, 5-1-5 Kashiwa-no-Ha, Kashiwa, Chiba, 277-8583, Japan}\\[2mm]
\noindent{\textbf{Email}: \href{mailto:pietro.caradonna@ipmu.jp}{\color{blue}{\smash{pietro.caradonna@ipmu.jp}}}}\\[4mm]
{\noindent{\bf Keywords:}{~Entanglement, Compton scattering, Bell States, Compton cameras, Compton PET}
 \lin{5}{7}

{ 
{\noindent{\large\textbf{Abstract}}{\\
\textup{This theoretical research aims to examine areas of the Compton cross section of entangled annihilation photons for the purpose of testing for possible break down of theory, which could have consequences for predicted optimal capabilities of Compton PET systems.We provide maps of the cross section for entangled annihilation photons for experimental verification.We introduce a strategy to derive cross sections in a relatively straight forward manner for the Compton scattering of a	hypothetical separable, mixed and entangled states. To understand the effect that entanglement has on the cross section for annihilation photons, we derive the cross section so that it is expressed in terms of the cross section of a hypothetical separable state and of a hypothetical forbidden maximally entangled
state.We find lobe-like structures in the cross section which are regions where entanglement has the greatest effect.We also find that mixed states do not reproduce the cross section for annihilation photons, contrary to a recent investigation which reported otherwise.We review the motivation and method of the most precise Compton scattering experiment for annihilation photons, in order to	resolve conflicting reports regarding the extent to which the cross section itself has been	experimentally verified.}}}
\\
  
\renewcommand{\headrulewidth}{0.5pt}
\renewcommand{\footrulewidth}{0pt}
 
 \pagestyle{fancy}
 \fancyfoot{}
 \fancyhead{} 
 \fancyhf{}
 \fancyhead[RO]{\leavevmode \put(-90,0){P. Caradonna \textit{et al.}}}
 \fancyhead[LE]{\leavevmode \put(0,0){P. Caradonna \textit{et al.}}}
 \fancyfoot[C]{\leavevmode
 \put(-2.5,-30){\thepage}}
 
 \renewcommand{\headrule}{\hbox to\headwidth{\leaders\hrule height \headrulewidth\hfill}}
 
\lin{1}{1}
\sectionn{Introduction}
{ \fontfamily{times}\selectfont
 \noindent 
In present day Positron Emission Tomography (PET) imaging systems, suspected false coincidence events are removed using energy discrimination and correction techniques. With the advent of Compton camera technology~\cite{Todd1974, Takahashi2012, Watanabe2009, Frandes2010,Moskal2016}, it has been suggested that entanglement of the annihilation photons can be used as an additional discriminator to increase the accuracy of eliminating false coincidence events, thereby improving image quality~\cite{McNamara2014}. However, before entanglement can be used, fundamental issues relating to kinematic outcomes of Compton scattering of annihilation photons must be resolved.
 
These kinematic outcomes are described by the Compton collision cross section, and an understanding of the capabilities of Compton PET systems requires an accurate estimate of the theoretical cross section. Previous experiments to measure the value of the anisotropy of the cross section yielded indirect confirmation of the predicted value, because the measured value of the anisotropy was inferred after the data were geometrically corrected~\cite{Wu1950,Langhoff1960,Kasday1971,Faraci1974,Kasday1975,Wilson1976,Bruno1977,Bertolini1981}. As far as this investigation is aware, the theoretical cross section has yet to be verified experimentally.

Verification of the cross section is also an issue of fundamental physics. Bohm and Aharonov~\cite{Bohm1957,Bohm1976} were first to recognize that the relative polarization correlations between annihilation photons was an example of the kind of entanglement discussed by Einstein, Podolsky and Rosen~\cite{Einstein1935}, and that the consequences of this entanglement can be observed in Compton scattering experiments. However, the results of these experiments are based on the assumption that the theoretical cross section is correct~\cite{Kasday1971,Kasday1975}.

Recently, Hiesmayr and Moskal argued that a certain mixed state photons which do not need to originate from the same positron-electron annihilation event gives the same Compton collision cross section as annihilation photons~\cite{Hiesmayr2019}. If true, this would limit the accuracy of Compton PET and would necessitate a reinterpretation of the results of past experiments testing the Bohm-Aharonov hypothesis.Wederive the cross section for that mixed state and compare our findings to previous results. As the level of experimental
verification of the cross section for annihilation photons is debated~\cite{Hiesmayr2019,Duarte2012}, we analyse previous reports to
resolve these inconsistencies.

\sectionn{Matrix representation of the Klein-Nishina formula}
\label{sec:Section2}
{ \fontfamily{times}\selectfont
	\noindent
We derive Compton collision cross sections for Bell state photons and mixed states using a matrix representation of the Klein-Nishina formula, a representation which was pioneered by Wightman~\cite{Wightman1948} and Fano~\cite{Fano1949} and thereafter by McMaster~\cite{McMaster1961} who demonstrated its versatility by deriving various cross sections in a relatively straight-forward and intuitive way. 

Herein, the photon polarization state is defined with respect to a linear polarization basis, and we adopt the source view convention and a right-handed coordinate system. The linear basis is a useful basis to represent the polarization state of the incoming and scattered photon. In this basis, the matrix representation of the Klein-Nishina formula $T(\theta; E_{o})$ is sufficiently described by the upper left $3\times3$ sub-matrix of the Fano $4\times4$ matrix given in~\cite{McMaster1961} such that 
\begin{equation}
T(\theta;E_{o})=\frac{r_{o}^{2}}{2}\left(\frac{E}{E_{o}}\right)^{2}\begin{bmatrix}1+\cos^{2}\theta+(E_{o}-E)(1-\cos\theta) & \sin^{2}\theta & 0 \\ \sin^{2}\theta & 1+\cos^{2}\theta & 0 \\  0 & 0 & 2 \cos\theta\\\end{bmatrix},
\label{eqn:Eqn1}
\end{equation}
where the Fano matrix $T(\theta;E_{o})$ is parametrized by the incident photon energy $E_{o}$, and the scattering angle $\theta$ subtends the incident and outgoing/scattered photon trajectories, $r_{o}\approx 2.18\times 10^{-15}$m is the classical electron radius, and $E$ is the energy of the outgoing photon. Equation (\ref{eqn:Eqn1}) considers the case in which the initial and final states are pure states, and the matrix elements have been summed over the final electron spin states. The quantities  $\theta$, $E$ and $E_{o}$ are related through the Compton formula 
\begin{equation}
E=\frac{E_{o}}{1+E_{o}(1-\cos\theta)},
\label{eqn:Eqn2}
\end{equation}
such that $E_{o}$ and $E$ are unitless quantities normalized using $mc^{2}$. For example, photons with an incident energy of $E_{o}=1$ correspond to energies of 511 keV ($mc^{2}$). 

\begin{figure}[t]
	\begin{center}
		\includegraphics[scale=0.5]{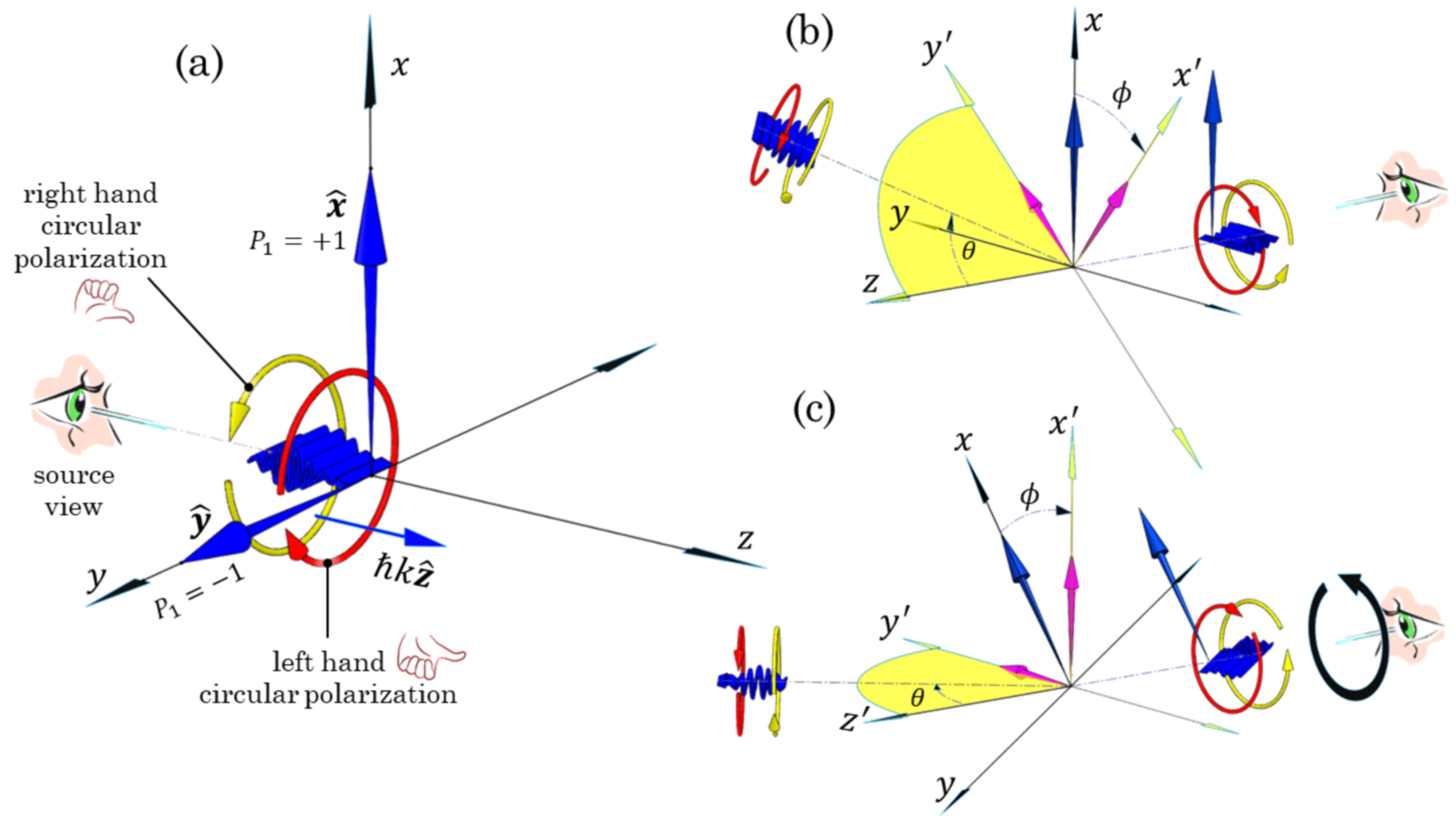}
		\caption{(a): A visualization of the Stokes parameters $P_{1}=\pm 1$ and their correspondence with the unit vectors $\xhat$ and $\yhat$, each of which indicate the plane of vibration of the classical electric field. The blue wave packet with momentum $\hbar k\zhat$ which has a right- and left-handed arcs, hovering around it, is used to crudely represent linearly polarized photon in a pure state which are in a superposition of right- and left- handed circular polarization states (Refer to Equation (\ref{eqn:Eqn7a})) (b): Shows a vertically polarized photon which is defined in the unprimed coordinate system and which scatters through an angle $\theta$ and $\phi$. (c): Shows the coordinate system rotated about the direction of propagation by an azimuthal angle $\phi$ into the primed coordinate system, which is analogous to applying a rotation matrix $M(\phi)$ (Equation (\ref{eqn:Eqn5})) which transforms the Stokes parameters from one coordinate system to another system. In the rotated frame, the scattering plane is defined in the $y^{'}$-$z^{'}$ plane.} 
		\label{fig:Fig1}
		\vspace{-10pt}
	\end{center}
\end{figure}

The Fano framework describes an incoming photon polarization state using a 4 component Stokes vector $\ket{S}$ (reference~\cite{McMaster1961} for more detail) and in its most general case is represented as
\begin{equation}
\ket{S}=\begin{bmatrix}I \\ P_{1}\\P_{2}\\P_{3}\end{bmatrix}=\begin{bmatrix}\bra{J} I_{2}\ket{J} \\ \bra{J} \sigma_{z}\ket{J} \\ \bra{J} \sigma_{x}\ket{J}\\ \bra{J} \sigma_{y}\ket{J}\end{bmatrix}=\begin{bmatrix}\abs{a_{1}}^{2}+\abs{a_{2}}^{2}\\ \abs{a_{1}}^2-\abs{a_{2}}^2\\ a_{1}a_{2}^{*}+a_{2}a_{1}^{*}\\ i(a_{1}a_{2}^{*}-a_{2}a_{1}^{*})\end{bmatrix},\quad\quad\ket{J}=\begin{bmatrix}a_{1}\\a_{2}\end{bmatrix},
\label{eqn:Eqn3}
\end{equation}
where $\ket{J}$ stands for a general Jones state vector with probability amplitudes $a_{1}$ and $a_{2}$, where $a_{1}, a_{2}\in\mathbb{C}$. Further detail regarding $P_{2}$ and $P_{3}$ is found in~\cite{McMaster1961}. Each component represents an observable quantity which are defined as the expectation value of the $2\times2$ unit matrix $I_{2}$ and the Pauli spin matrices $\{\sigma_{x}, \sigma_{y}, \sigma_{z}\}$. When working in the linear basis, the component $P_{3}$ can be omitted, in which case the Stokes vectors of interest for this work are
\begin{equation}
\mathbf{\hat{x}}\equiv\ket{+}=\begin{bmatrix}1\\+1\\0\end{bmatrix},\quad\mathbf{\hat{y}}\equiv\ket{-}=\begin{bmatrix}1\\-1\\0\end{bmatrix},\quad\mbox{and}\quad\ket{I}=\frac{1}{2}\ket{+}+\frac{1}{2}\ket{-}=\begin{bmatrix}1\\0\\0\end{bmatrix},
\label{eqn:Eqn4}
\end{equation}
where $\ket{+}$ and $\ket{-}$ label the Stokes vector for vertically and horizontally polarized photons and correspond to the unit vectors $\mathbf{\hat{x}}$ and $\mathbf{\hat{y}}$, respectively, as illustrated in Figure \ref{fig:Fig1}}(a). The Stokes vector labelled $\ket{I}$ represents a normalized beam of unpolarized photons.

The Stokes parameters are also dependent on the choice of coordinate systems. The Fano matrix method normally requires a rotation matrix $M(\phi)$ which maps the Stokes parameters given in Equation (\ref{eqn:Eqn4}) to the Stokes parameters in an another system and is given by
\begin{equation}
M(\phi)=
\begin{bmatrix}1&0&0\\0&\cos2\phi&\sin2\phi\\0&-\sin2\phi&\cos2\phi\end{bmatrix}.
\label{eqn:Eqn5}
\end{equation}
With respect to the source view, figures \ref{fig:Fig1}(b-c) shows that a counterclockwise rotation about the incident photon trajectory by the azimuthal angle $\phi$ corresponds to a clockwise rotation of the unprimed coordinate system, which in Stokes space is a rotation about the axis represented by $P_{3}$, and where we make use of the following properties
\begin{equation}
M\left(\phi\pm\frac{\pi}{2}\right)\ket{-}=M\left(\phi\right)\ket{+},\quad M\left(\phi\pm\frac{\pi}{2}\right)\ket{+}=M\left(\phi\right)\ket{-},\quad M\left(\phi\right)\ket{I}=\ket{I}.
\label{eqn:Eqn6}
\end{equation} 

An experiment which yields linear polarized light is a result of detecting an even amount of right and left hand circularly polarized photons each of which carry angular momentum of $+\hbar\mathbf{\hat{z}}$ and $-\hbar\mathbf{\hat{z}}$, with respect to the projection of the angular momentum onto the axis of propagation. The basis vectors for right and left hand circularly polarized photons are labeled by the Jones vectors $\ket{R}$ and $\ket{L}$, respectively. Since this basis carries the added information of the intrinsic spin of a photon, we consider this basis more fundamental than compared to the vertical ($\ket{\uparrow}$) and horizontal ($\ket{\rightarrow}$) Jones vectors, since an experiment which measures linear polarization yields no information of the intrinsic angular momentum. The linear basis can be expanded in terms of $\ket{R}$ and $\ket{L}$ such that
\begin{subequations}
	\begin{equation}
	\ket{\uparrow}=\begin{bmatrix}1\\0\end{bmatrix}=\frac{1}{\sqrt{2}} \ket{R}+\frac{1}{\sqrt{2}} \ket{L},\quad\ket{\rightarrow}=\begin{bmatrix}0\\1\end{bmatrix}=\frac{i}{\sqrt{2}} \ket{R}-\frac{i}{\sqrt{2}}\ket{L},
		\label{eqn:Eqn7a}
	\end{equation}
	where
	\begin{equation}
	\ket{\mbox{R}}=\frac{1}{\sqrt{2}}\begin{bmatrix}1\\-i\end{bmatrix},\quad\ket{\mbox{L}}=\frac{1}{\sqrt{2}}\begin{bmatrix}1\\i\end{bmatrix}.
	\label{eqn:Eqn7b}
	\end{equation}    
\end{subequations}
The states given in equation (\ref{eqn:Eqn7a}) show that a linearly polarized photon is a photon which is in an indefinite state of angular momentum. The sketch of the blue wave packet shown in figures \ref{fig:Fig1}} (a-c) which has the orbiting right and left hand circular arcs is used as a visual aid to emphasis the potentiality of a photon to collapse in either state $\ket{R}$ or $\ket{L}$ through some appropriate experimental procedure. Throughout this article, both $\ket{\uparrow}$ and $\ket{\rightarrow}$ will be visualized in this way. To distinguish between the two linear states, we attach a vertical or horizontal arrow indicating which of the two linear states we are referring to.  
 
\newpage
\subsectionn{Rules for deriving cross sections for mixed and pure states}
{ \fontfamily{times}\selectfont
	\noindent
The objective is  apply the Fano matrix to derive probabilities distributions for Compton scattering by pairs of photons in Bell states and mix states and show the relationship between these distributions. The distributions are derived using the established rules of quantum mechanics which were presented in a form more familiar to physicists by Furry~\cite{Furry1936}. In particular, in the case when the probability $P_{pure}$ represents the probability of a pure state $\ket{\Psi}$ that passes a test for being in the state $\ket{\Phi}$ is
\begin{equation}
P_{pure}=\abs{\bra{\Phi}\ket{\Psi}}^{2}.
\label{eqn:Eqn8}
\end{equation}
Whereas, for a mixture of states when we only know the probabilities $\omega_{i}$ of the system being in the states $\ket{\Psi_{i}}$, the probability $P_{mix}$ of a mixed state that passes a test for being in the state $\ket{\Phi}$ the probability in question is
\begin{equation}
P_{mix}=\sum_{i}w_{i}\abs{\bra{\Phi}\ket{\Psi_{i}}}^{2}.
\label{eqn:Eqn9}
\end{equation}

In a photon counting experiment, the probability distribution for Compton scattering of a photon consists of a causal relation between two events, namely an event $n_{s}$ which announces to the experimenter that an incoming photon has Compton scattered, and a subsequent event $n\{\theta,\phi\}$ which announces to the experimenter that the outgoing photon has interacted with a detector positioned at coordinates $(r,\theta, \phi)$ relative to the scattering site. Thus, the probability distributions for Compton scattering are conditional probabilities of the form $P(n\{\theta,\phi\}|n_{s})$, which is taken to mean; {\it{Given that a Compton scattering event $n_{s}$ has occurred, with what probability $P$ will an event $n\{\theta,\phi\}$ occur in a detector positioned at coordinate $(r,\theta, \phi)$?}} 

The conditional probabilities for Compton scattering are proportional to the differential collision cross sections. For an  unpolarized beam with incident energy $E_{o}$, which comprises, for example, of pure state photons is represented by the Stokes vector $\ket{I}$. The procedure of using the Fano method to derive differential cross sections is to apply the necessary vectors and matrices of Section (\ref{sec:Section2}) from right to left, beginning with the Stokes vector representation of the state under consideration, in this case $\ket{I}$. To a Stokes vector, one would typically apply the rotation matrix $M(\phi)$ (\ref{eqn:Eqn5}), then act on this transformed vector with the Fano matrix (\ref{eqn:Eqn1}). However, equation (\ref{eqn:Eqn6}) implied that an unpolarized state $\ket{I}$ is invariant under rotation, i.e., $M(\phi)\ket{I}=\ket{I}$. Therefore it is sufficient to apply the Fano matrix to the state $\ket{I}\mapsto T(\theta;E_{o})\ket{I}$. The final step is to apply the Stokes bra vector $\bra{I}$, which is the Stokes vector characterizing of a photon counter such as a scintillation detector. The differential cross section for a linearly unpolarized beam is    
\begin{equation}
\frac{d\sigma}{d\Omega}=\bra{I}T(\theta;E_{o})\ket{I}=\frac{r_{o}^{2}}{2}\left(\frac{E}{E_{o}}\right)^{2}\left(\frac{E}{E_{o}}+\frac{E_{o}}{E}-\sin^{2}\theta\right),
\label{eqn:Eqn10}
\end{equation}
which, when evaluated, gives the well known Klein-Nishina Compton collision cross section in the case for a polarization-insensitive detector. Integrating Equation (\ref{eqn:Eqn10}) over all angles for the case of annihilation photons ($E_{o}=1$) gives for the total collision cross section
\begin{equation}
\sigma=\int_{\theta=0}^{\pi}d\theta\int_{\phi=0}^{2\pi}d\phi\bra{I}T(\theta;1)\ket{I}\sin\theta=\frac{\pi r_{o}^{2}}{9}(40-27\ln3),\quad E_{o}=1\quad(511 \mbox{keV}).
\label{eqn:Eqn11}
\end{equation}
The probability of detecting an unpolarized photon beam (equal mix of right and left circularly polarized photons) using a polarization-insensitive detector is
\begin{equation}
P(n\{\theta,\phi\}|n_{s})\approxeq\frac{\Delta\Omega}{\sigma}\bra{I}T(\theta;1)\ket{I}=\frac{\Delta\Omega}{\sigma}\left[\frac{r_{o}^{2}}{2}\frac{(1-\cos\theta)^{3}+2}{(2-\cos\theta)^{3}}\right],\quad\mbox{for}\quad E_{o}=1,
\label{eqn:Eqn12}
\end{equation}
where $\Delta\Omega=\Delta\theta\Delta\phi\sin\theta$ represents the solid angle into which the photon scatters into, and that $\Delta\Omega$ is sufficiently small such that the differential cross section is approximately constant over this interval about the angle $\theta$ and $\phi$. As can be seen from equation (\ref{eqn:Eqn12}), in the case of an unpolarized photon beam, the probability of detecting an event $n\{\theta,\phi\}$ is independent of the azimuthal angle $\phi$. 

\sectionn{Compton scattering of entangled states}
{ \fontfamily{times}\selectfont
	\noindent
Parapositronium which is at rest annihilates into predominately two entangled photons (labeled $\gamma_{1i}$ and $\gamma_{2i}$) that propagate in the $\mathbf{\hat{z}}$ direction, respectively.Due to parity, angular momentum conservation~\cite{Yang1950} and Bose symmetry, the entangled state is
\begin{equation}
\ket{\Phi^{-}_{c}}=\frac{1}{\sqrt{2}}\Big[\ket{R_{1i}}\otimes\ket{R_{2i}}-\ket{L_{1i}}\otimes\ket{L_{2i}}\Big]\longmapsto\ket{\Psi^{+}_{l}}=\frac{-i}{\sqrt{2}}\Big[\ket{\uparrow_{1i}}\otimes\ket{\rightarrow_{2i}}+\ket{\rightarrow_{1i}}\otimes\ket{\uparrow_{2i}}\Big],
\label{eqn:Eqn13}
\end{equation}
where $\ket{\Phi^{-}_{c}}$ and $\ket{\Psi^{+}_{l}}$ are particular cases of the maximally entangled Bell states, each of which is an equivalent representation of the aforementioned process, and the subscripts $c$ and $l$ denote {\it{``circular"}} and {\it{``linear"}} polarization basis, respectively. 
\begin{figure*}[t]
	\includegraphics[scale=0.5]{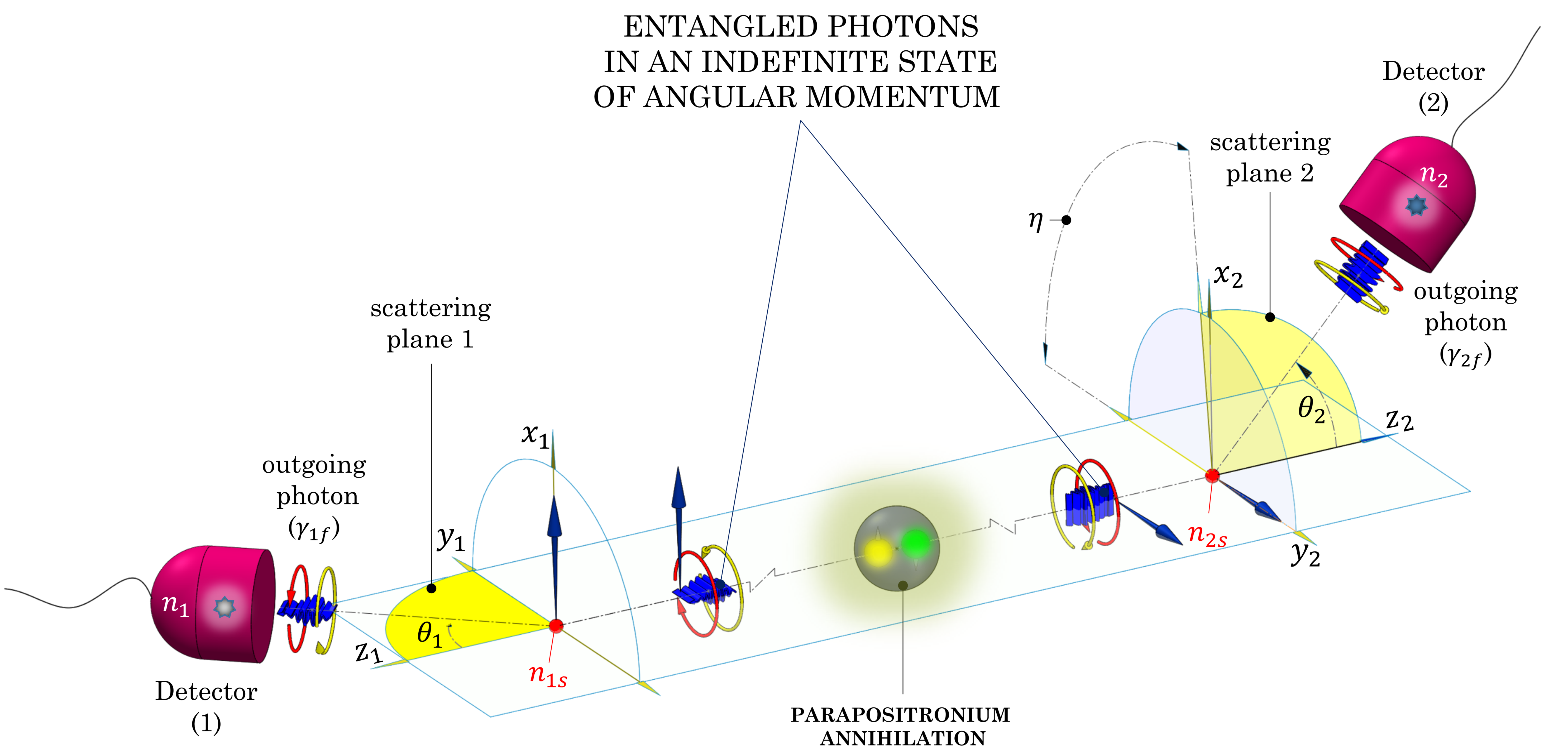}
	\caption{The nomenclature for Compton scattering by a pair of entangled 511 keV photons in the state corresponding to $\ket{\uparrow_{1i}}\otimes\ket{\rightarrow_{2i}}$. The diagramdepicts the annihilation in the rest frame of parapositronium, and produce photons in an indefinite state of angular momentum. In this frame, the photons travel along a common straight line in opposite directions, and after some elapsed time each undergo a Compton scattering event with a stationary electron at the origin of their respective coordinate system. The trajectories of each scattered photon $\gamma_{1f}$ and $\gamma_{2f}$ lie in the yellow shaded scattering plane, respectively. {\it(Note: Scattered electrons not shown.)}}  
	\label{fig:Fig2}
\end{figure*}

Using the state $\ket{\Psi^{+}_{l}}$, the Compton scattering of the state $\ket{\uparrow_{1i}}\otimes\ket{\rightarrow_{2i}}$  is visualized in figure (\ref{fig:Fig2}), where it is shown the angle $\eta$ defines the angular separation between the two scattering planes. The Fano matrix representation of the Compton collision cross section by the subsequent emission of Bell state photons by parapositronium annihilation can be represented by  
\begin{equation}
\frac{\partial^{2}\sigma^{(\Psi_{l}^{+})}}{\partial\Omega_{1}\partial\Omega_{2}}=\frac{1}{8}\sum\limits_{a=0}^{1}\bra{I}T(\theta_{1};1)M\Big(\frac{a\pi}{2}\Big)\ket{+}\bra{I}T(\theta_{2};1)M\Big(\eta+\frac{a\pi}{2}\Big)\ket{-},
\label{eqn:Eqn14}
\end{equation}
where the sum in equation (\ref{eqn:Eqn14}) is over two possible final states, and the factor of $1/8$ is the averaging over the 8 degrees of freedom in the initial states can scatter off an incident electron, with respect to the scattering plane of $\gamma_{1f}$. The differential cross section of equation (\ref{eqn:Eqn14}) is equivalent in structure to the Pryce-Ward equation given in~\cite{Pryce1947} and may be expressed in the following form
\begin{subequations}
	\begin{equation}
	\frac{\partial^{2}\sigma^{(\Psi_{l}^{+})}}{\partial\Omega_{1}\partial\Omega_{2}}=\frac{1}{4}\bra{I}T(\theta_{1};1)\ket{I}\bra{I}T(\theta_{2};1)\ket{I}\bigg[1-m(\theta_{1})m(\theta_{2})\cos 2\eta\bigg],
	\label{eqn:Eqn15a}
	\end{equation}
	where the term $\bra{I}T(\theta_{1};1)\ket{I}\bra{I}T(\theta_{2};1)\ket{I}$ is the product of the Klein-Nishina cross section given in equation (\ref{eqn:Eqn10}), evaluated for photons with incident energy of 511 keV, and is proportional to the probability of Compton scattering of a pair of mutually independent photons propagating in the same back-to-back configuration, and where
	\begin{equation}
	m(\theta)=\frac{(2-\cos\theta)\sin^{2}\theta}{(1-\cos\theta)^{3}+2}.
		\label{eqn:Eqn15b}
	\end{equation}
\end{subequations}	
The Compton scattering of Bell state photons involves four causally related events. Specifically, the event $N_{ss}=n_{1s}\cap n_{2s}$ which is the result of the Compton scattering by each of the incoming photons, and two subsequent events $n_{1}\{\theta_{1},\phi_{1}\}$ and $n_{2}\{\theta_{2},\phi_{2}\}$ at detectors (1) and (2) which are positioned at coordinates $(r_{1},\theta_{1}, \phi_{1})$ and $(r_{2},\theta_{2}, \phi_{2})$, respectively. Therefore, the conditional probability of Compton scattering of Bell state photons is of the form $P^{(\Psi_{l}^{+})}(n_{1}\{\theta_{1},\phi_{1}\},n_{2}\{\theta_{2},\phi_{2}\}|N_{ss})$. It follows then that 
\begin{equation}
P^{(\Psi_{l}^{+})}(n_{1}\{\theta_{1}\},n_{2}\{\theta_{2},\eta\}|N_{ss})\approxeq\frac{\Delta\Omega_{1}\Delta\Omega_{2}}{\sigma_{(\gamma_{1},\gamma_{2})}}\frac{\partial^{2}\sigma^{(\Psi_{l}^{+})}}{\partial\Omega_{1}\partial\Omega_{2}}=\frac{4\Delta\Omega_{1}\Delta\Omega_{2}}{\sigma^{2}}\frac{\partial^{2}\sigma^{(\Psi_{l}^{+})}}{\partial\Omega_{1}\partial\Omega_{2}},
\label{eqn:Eqn16}
\end{equation}
where
\begin{equation}
\sigma_{(\gamma_{1}, \gamma_{2})}=\iint\limits_{\partial\Omega_{1}\partial\Omega_{2}}\partial^{2}\sigma^{(\Psi_{l}^{+})}=\left[\frac{\pi r_{o}^{2}}{18}\left(40-27\ln3\right)\right]^{2}=\frac{1}{4}\sigma^{2}.
\label{eqn:Eqn17}
\end{equation} 


We now consider some hypothetical mechanism that produces a pair of maximally entangled photons represented by the state 
\begin{equation}
\ket{\Psi_{c}^{+}}=\frac{1}{\sqrt{2}}\ket{R_{1i}}\otimes\ket{L_{2i}}+\frac{1}{\sqrt{2}}\ket{L_{1i}}\otimes\ket{R_{2i}}\longmapsto\ket{\Phi_{l}^{+}}=\frac{1}{\sqrt{2}}\ket{\uparrow_{1i}}\otimes\ket{\uparrow_{2i}}+\frac{1}{\sqrt{2}}\ket{\rightarrow_{1i}}\otimes\ket{\rightarrow_{2i}},\\
\label{eqn:Eqn18}
\end{equation}
where the state $\ket{R_{1i}}\otimes\ket{L_{2i}}$ and $\ket{L_{1i}}\otimes\ket{R_{2i}}$ carry a total of $\pm 2\hbar\mathbf{\hat{z}_{2}}$ of angular momentum, respectively. Figure (\ref{fig:Fig3}) visualizes the geometry for Compton scattering by the state $\ket{\uparrow_{1i}}\otimes\ket{\uparrow_{2i}}$, which is similar to the geometry given in figure (\ref{fig:Fig2}). The differential Compton cross section for the state given in equation (\ref{eqn:Eqn18}) is  
\begin{figure*}[t]
		\includegraphics[scale=0.5]{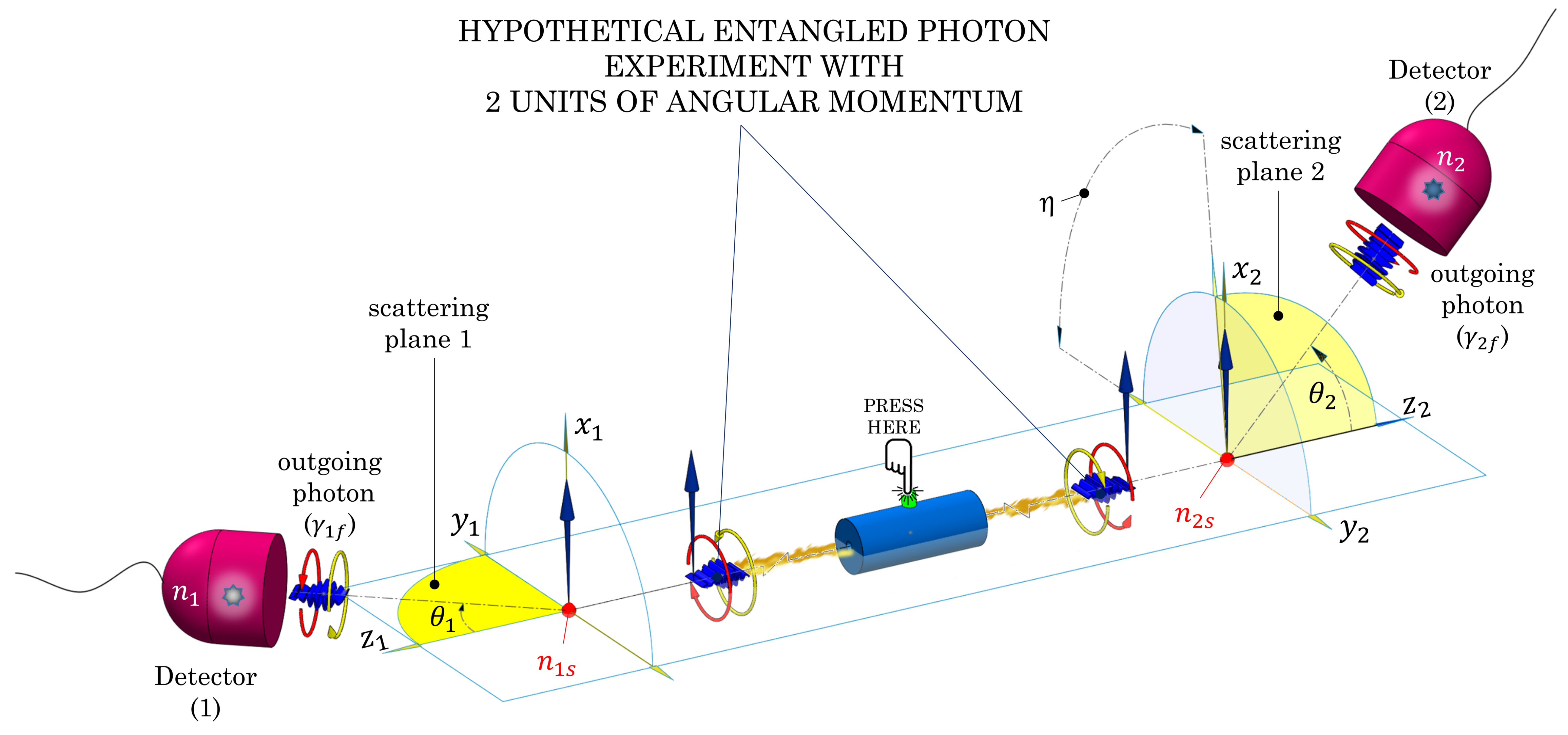}
	\caption{The nomenclature for Compton scattering by a hypothetical pair of 511 keV photons  in the state corresponding to $\ket{\uparrow_{1i}}\otimes\ket{\uparrow_{2i}}$, of which became maximally entangled through some mechanism. In this frame, the photons travel  in opposite directions along the $z-$axis, and carries a total of $2\hbar\mathbf{\hat{z}_{2}}$ of angular momentum for $\ket{R_{1i}}\otimes\ket{L_{2i}}$ and $-2\hbar\mathbf{\hat{z}_{2}}$ for $\ket{L_{1i}}\otimes\ket{R_{2i}}$. After some elapsed time, each photon Compton scatters at the origin of their respective coordinate systems.} 
	\label{fig:Fig3}
\end{figure*}

\begin{subequations}
	\begin{equation}
	\begin{aligned}
	&\frac{\partial^{2}\sigma^{(\Phi_{l}^{+})}}{\partial\Omega_{1}\partial\Omega_{2}}=\frac{1}{8}\sum_{a=0}^{1}\bra{I}T(\theta_{1};1)M\left(\frac{a\pi}{2}\right)\ket{+}\bra{I}T(\theta_{2};1)M\left(\eta+\frac{a\pi}{2}\right)\ket{+},
	\end{aligned}
	\label{eqn:Eqn19a}
	\end{equation}

	\begin{equation}
	\implies\frac{\partial^{2}\sigma^{(\Phi_{l}^{+})}}{\partial\Omega_{1}\partial\Omega_{2}}=\frac{1}{4}\bra{I}T(\theta_{1};1)\ket{I}\bra{I}T(\theta_{2};1)\ket{I}\bigg[1+m(\theta_{1})m(\theta_{2})\cos 2\eta\bigg],
	\label{eqn:Eqn19b}
	\end{equation}
\end{subequations}
which implies	
\begin{equation}
P^{(\Phi_{l}^{+})}(n_{1}\{\theta_{1}\},n_{2}\{\theta_{2},\eta\}|N_{ss})\approxeq\frac{4\Delta\Omega_{1}\Delta\Omega_{2}}{\sigma^{2}}\frac{\partial^{2}\sigma^{(\Phi_{l}^{+})}}{\partial\Omega_{1}\partial\Omega_{2}}.
\label{eqn:Eqn20}
\end{equation}

One can find a relationship between the differential cross sections for the Bell states given by equations (\ref{eqn:Eqn14}) and (\ref{eqn:Eqn19a}) by considering a hypothetical ensemble of the state $\ket{\Psi^{+}_{l}}$ (\ref{eqn:Eqn13}) and $\ket{\Phi^{+}_{l}}$ (\ref{eqn:Eqn18}) represented by the following density matrix $\rho_{mix}^{bell}$ such that 
\begin{equation}
\rho_{mix}^{bell}=\frac{1}{2}\ket{\Psi_{l}^{+}}\bra{\Psi_{l}^{+}}+\frac{1}{2}\ket{\Phi_{l}^{+}}\bra{\Phi_{l}^{+}},
\label{eqn:Eqn21}
\end{equation}
where the probability of finding an individual system is $1/2$. 

Applying equation (\ref{eqn:Eqn9}) to this mixed state gives for the probability of Compton scattering
\begin{equation}
\begin{aligned}
P_{mix}^{bell}(n_{1}\{\theta_{1}\},n_{2}\{\theta_{2},\eta\}|N_{ss})&=\frac{1}{2}P^{(\Psi_{l}^{+})}(n_{1}\{\theta_{1}\},n_{2}\{\theta_{2},\eta\}|N_{ss})+\frac{1}{2}P^{(\Phi_{l}^{+})}(n_{1}\{\theta_{1}\},n_{2}\{\theta_{2},\eta\}|N_{ss})\\
&\approxeq\frac{\Delta\Omega_{1}\Delta\Omega_{2}}{\sigma^{2}}\bra{I}T(\theta_{1};1)\ket{I}\bra{I}T(\theta_{2};1)\ket{I}.
\end{aligned}
\label{eqn:Eqn22}
\end{equation}
Rearranging equation (\ref{eqn:Eqn22}) with resepect to $P^{(\Psi_{l}^{+})}(n_{1}\{\theta_{1}\},n_{2}\{\theta_{2},\eta\}|N_{ss})$ gives
\begin{equation}
P^{(\Psi_{l}^{+})}(n_{1}\{\theta_{1}\},n_{2}\{\theta_{2},\eta\}|N_{ss})\approxeq\frac{2\Delta\Omega_{1}\Delta\Omega_{2}}{\sigma^{2}}\bra{I}T(\theta_{1};1)\ket{I}\bra{I}T(\theta_{2};1)\ket{I}-\frac{4\Delta\Omega_{1}\Delta\Omega_{2}}{\sigma^{2}}\frac{\partial^{2}\sigma^{(\Phi_{l}^{+})}}{\partial\Omega_{1}\partial\Omega_{2}}.
\label{eqn:Eqn23}
\end{equation}

In terms of the differential cross section for the Bell states $\ket{\Psi^{+}_{l}}$ (\ref{eqn:Eqn15a}) and $\ket{\Phi_{l}^{+}}$ (\ref{eqn:Eqn19b}), equation (\ref{eqn:Eqn22}), we find
\begin{equation}
\frac{\partial^{2}\sigma^{(\Psi_{l}^{+})}}{\partial\Omega_{1}\partial\Omega_{2}}=\frac{1}{2}\bra{I}T(\theta_{1};1)\ket{I}\bra{I}T(\theta_{2};1)\ket{I}-\frac{\partial^{2}\sigma^{(\Phi_{l}^{+})}}{\partial\Omega_{1}\partial\Omega_{2}}.
\label{eqn:Eqn24}
\end{equation}
When the differential cross section for Compton scattering by a pair of photons emitted in parapositronium annihilation is expressed in this form, the cross section is seen as an interplay between the Compton scattering of a pair of mutually independent photons represented by the term $\bra{I}T(\theta_{1};1)\ket{I}\bra{I}T(\theta_{2};1)\ket{I}$ that, as we shall show, possesses eigenfunctions of the emitted photons which are either individually allowed or forbidden in parapositronium annihilation and the term which describes a maximally entangled Bell state which has eigenfunctions which are strictly forbidden in connection to parapositronium annihilation.

\sectionn{Structures in the Compton cross section of annihilation photons}
{ \fontfamily{times}\selectfont
	\noindent
Figure (\ref{fig:Fig4}) is the four dimensional probability density function given by equation (\ref{eqn:Eqn23}). The color coding provides information about the probability of $\gamma_{1i}$ and $\gamma_{2i}$ Compton scattering through the angles $\theta_{1}$ and $\theta_{2}$, respectively, when the scattering planes are separated by an azimuthal angle of $\eta$. 
	
To visualize its interior region, the density function (the bulk) is sliced in half along $\theta_{1}=90^{\circ}$. Referring to Figure (\ref{fig:Fig4}), the two halves are labeled Left Sector (LS) and Right Sector (RS) defined within an angular range of $[$$($$\theta_{1}$, $0^{\circ}$, $90^{\circ}$$)$, $($$\theta_{2}$, $0^{\circ}$, $180^{\circ}$$)$, $($$\eta$, $0^{\circ}$, $360^{\circ}$$)$$]_{\text{LS}}$ and $[$$($$\theta_{1}$, $90^{\circ}$, $180^{\circ}$$)$, $($$\theta_{2}$, $0^{\circ}$, $180^{\circ}$$)$, $($$\eta$, $0^{\circ}$, $360^{\circ}$$)$$]_{\text{RS}}$, respectively. 
	
We refer to the center plot as the Interior Sector (IS) which has an angular range of $[$$($$\theta_{1}$, $75^{\circ}$, $115^{\circ})$, $(\theta_{2}$, $75^{\circ}$, $115^{\circ}$$)$, $($$\eta$, $0^{\circ}$, $360^{\circ}$$)$$]_{\text{IS}}$. The probability densities of the lobes which are made visible in the interior sector are around 10 times smaller relative to the immediate region that surrounds them. These are the regions where the probability distribution connected to the forbidden Bell state has the greatest effect on the cross section. When viewed from within the left or right sector, these lobes show up as faint white structures in the LS and RS are indicated by the dotted-outlined circles. It is for this reason the color scale in the interior sector has been enhanced to delineate these structures. Figure (\ref{fig:Fig5}) provides a closer examination of the central lobe mapped in terms of contours of constant percentage probability of Compton scattering. As far as we are aware, the topology of these structures are a unique signature of annihilation photons which have not been reported in the literature, and therefore have not undergone a direct empirical examination. 
	
\begin{figure*}[t]
\includegraphics[scale=0.5]{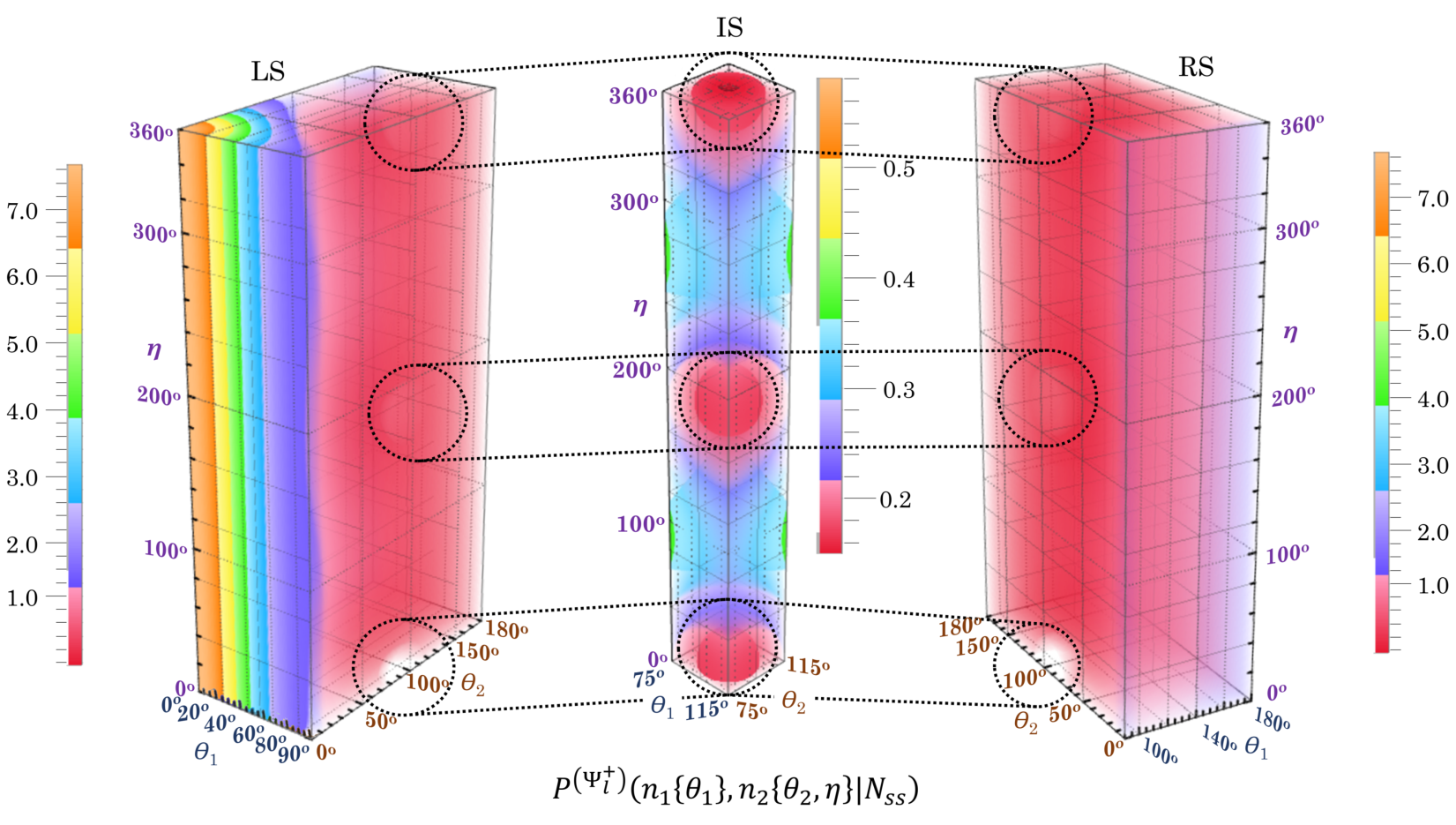}
\caption{A four-dimensional probability plot of the angular distribution of Compton scattered 511 keV photons. The plot is a visualization of  expressed as a percentage probability per unit of solid angle squared i.e., $P^{(\Psi_{l}^{+})}(n_{1}\{\theta_{1}\},n_{2}\{\theta_{2},\eta\}|N_{ss})$, equation (\ref{eqn:Eqn23}). The egg-shaped lobes shown in the inner sector (IS) have been re-scaled in order to delineate the structures. In the LS and RS region indicated by the dotted circles, the lobes appear as faint white regions. {\it{Note: $\Delta\Omega_{1}=\Delta\Omega_{2}=1$}}} 
\label{fig:Fig4}
\end{figure*}
	
\begin{figure}[t]
	\includegraphics[scale=0.3]{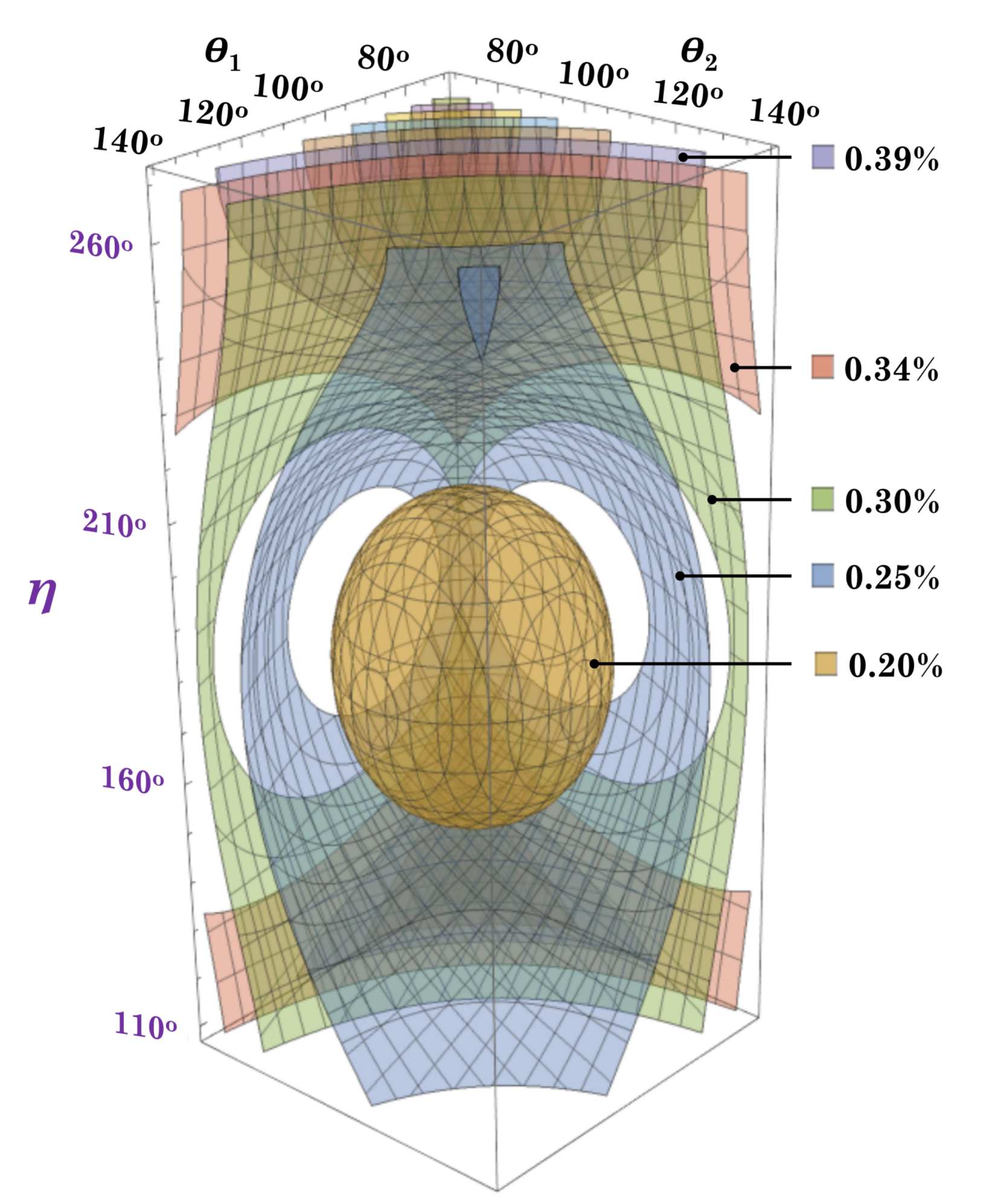}
	\caption{A closer view of a lobe structure in terms of contours of constant percentage probability given by $P^{(\Psi_{l}^{+})}(n_{1}\{\theta_{1}\},n_{2}\{\theta_{2},\eta\}|N_{ss})$, equation (\ref{eqn:Eqn15a}).} 
	\label{fig:Fig5}
	\vspace{-10pt}
\end{figure}
\newpage	
\sectionn{Discussion}
\subsectionn{On the uniqueness of the Compton cross section of annihilation photons}
{ \fontfamily{times}\selectfont
	\noindent
Of particular interest to this work is the mixed state       
\begin{equation}
\rho_{mix}=\frac{1}{2}\ket{\uparrow_{1i},\rightarrow_{2i}}\bra{\uparrow_{1i},\rightarrow_{2i}}+\frac{1}{2}\ket{\rightarrow_{1i},\uparrow_{2i}}\bra{\rightarrow_{1i},\uparrow_{2i}},
\label{eqn:Eqn25}
\end{equation}
as opposed to the density matrix for the pure state given in equation (\ref{eqn:Eqn13}) is
\begin{equation}
\rho^{(\Psi^{+}_{l})}=\rho_{mix}+\frac{1}{2}\bigg[\ket{\uparrow_{1i},\rightarrow_{2i}}\bra{\rightarrow_{1i},\uparrow_{2i}}+\ket{\rightarrow_{1i},\uparrow_{2i}}\bra{\uparrow_{1i},\rightarrow_{2i}}\bigg].
\label{eqn:Eqn26}
\end{equation}
This density matrix contains interference terms that are not present in equation (\ref{eqn:Eqn25}) and it is these additional terms that Bohm and Aharonov~\cite{Bohm1957} seized on to physically distinguish between these two states in Compton scattering experiments to explain the well-known EPR paradox~\cite{Einstein1935}.

In the case of the mixed state $\rho_{mix}$, which is illustrated in Figure (\ref{fig:Fig6}), the beam of photons entering either Detector (1) or (2) now contains an equal mix of the states $\ket{\uparrow}$ and $\ket{\rightarrow}$ whose cross sections is given by
\begin{equation}
\frac{1}{2}\bra{I}T(\theta_{a};1)M(\phi_{a})\ket{+}+\frac{1}{2}\bra{I}T(\theta_{a};1)M(\phi_{a})\ket{-}=\bra{I}T(\theta_{a};1)\ket{I}\quad a=1,2,
\label{eqn:Eqn27}
\end{equation}       
where we have used the property $1/2\ket{+}+1/2\ket{-}=\ket{I}$ and $M(\phi)\ket{I}=\ket{I}$, c.f. equations (\ref{eqn:Eqn4}) and (\ref{eqn:Eqn6}). Therefore, the joint probability of the mixed state $\rho_{mix}$ is proportional to 
\begin{equation}
P_{mix}(n_{1}\{\theta_{1}\},n_{2}\{\theta_{2},\eta\}|N_{ss})\propto \bra{I}T(\theta_{1};1)\ket{I}\bra{I}T(\theta_{2};1)\ket{I}.
\label{eqn:Eqn28}
\end{equation}
\begin{figure*}[t]
	\includegraphics[scale=0.5]{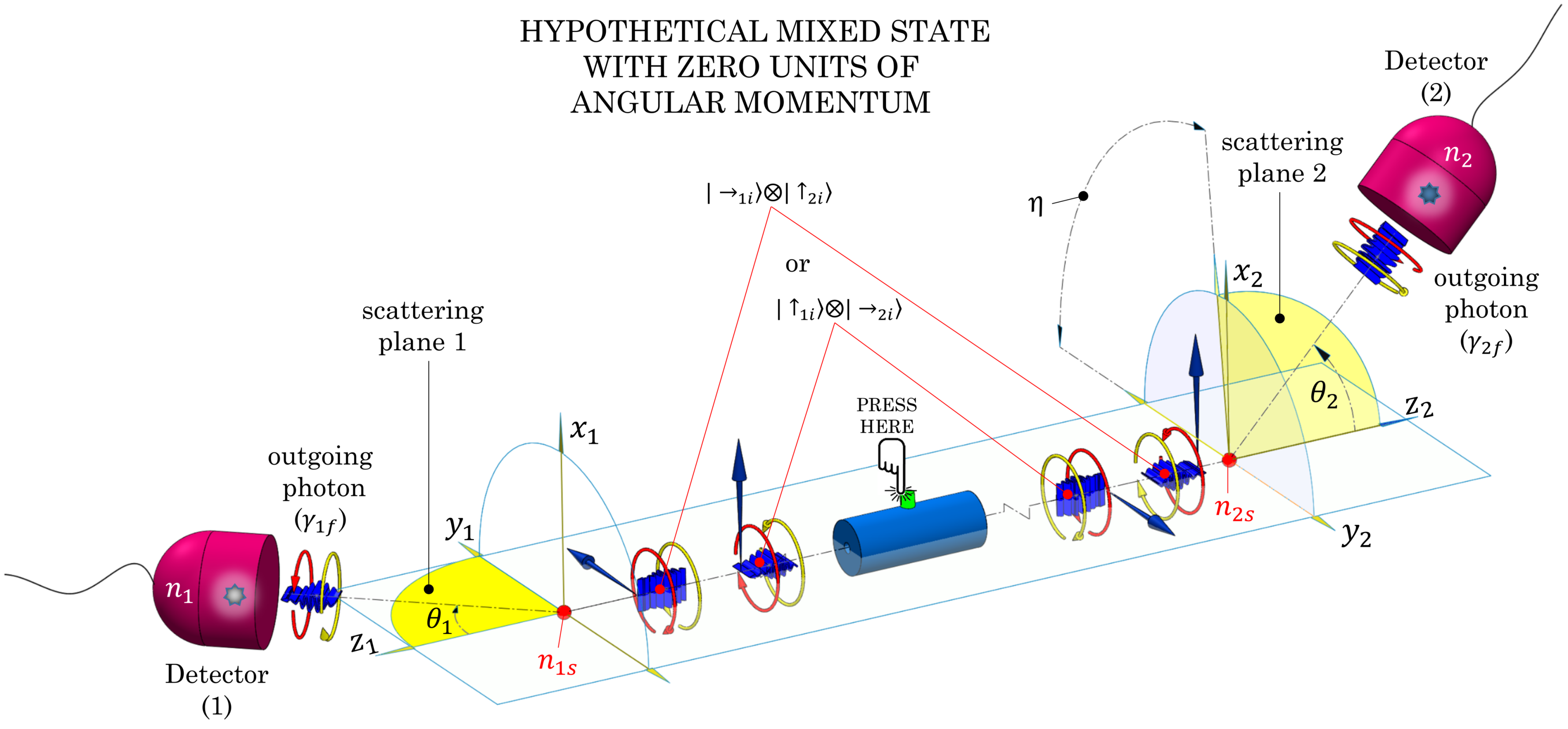}
	\caption{Compton scattering by the mixture of states given by state in (\ref{eqn:Eqn24}), which describes pairs of 511 keV photons in the state $\ket{\uparrow_{1i}}\otimes\ket{\rightarrow_{2i}}$ or $\ket{\rightarrow_{1i}}\otimes\ket{\uparrow_{2i}}$. The diagram depicts the creation of the mixed state by some unknown process. In this frame, the photons travel along a common straight line in opposite directions, and after some elapsed time each undergo a Compton scattering event with a stationary electron at the origin of their respective coordinate system. The trajectories of each scattered photon $\gamma_{1f}$ and $\gamma_{2f}$ lie in the yellow shaded scattering plane, respectively. {\it(Note: Scattered electrons not shown.)}} 
\label{fig:Fig6}
\end{figure*}
In Compton scattering experiments which measure the ratio $R=N_{\perp}/N_{\parallel}$ of counting rates (in which $\theta$ is fixed) in the two azimuthal planes corresponding to $\eta=0^{o}$ ($N_{\parallel}$) and $\eta=90^{o}$ ($N_{\perp}$), we find for the mixed state $\rho_{mix}$, equation (\ref{eqn:Eqn25}), that
\begin{equation}
R_{(\rho_{mix})}=\left(\frac{N_{\perp}}{N_{\parallel}}\right)_{\rho_{mix}}=\frac{\bra{I}T(\theta_{1};1)\ket{I}\bra{I}T(\theta_{2};1)\ket{I}}{\bra{I}T(\theta_{1};1)\ket{I}\bra{I}T(\theta_{2};1)\ket{I}}=1.
\label{eqn:Eqn29}
\end{equation}

In the case of the maximally entangled annihilation photons, refer to equation (\ref{eqn:Eqn13}), Pryce and Ward~\cite{Pryce1947}, and Snyder et al.~\cite{Snyder1948} have shown that the ratio for ideal geometries has a maxium value of $R=2.85$ for the Compton polar angles $\theta_{1}=\theta_{2}\approxeq81.67^{o}$. 
		
The solution of equation (\ref{eqn:Eqn29}) is consistent with that obtained by Bohm and Aharonov~\cite{Bohm1957}. Both our findings and that of Bohm and Aharonov are in formal disagreement with the recent investigation by Hiesmayr and Moskal~\cite{Hiesmayr2019}. Aside from the errors of missing $r_{o}/2$ factors in their differential cross sections, our findings and that of Bohm and Aharonov is different in a very important way in that Hiesmayr and Moskal concluded that the Compton collision cross section of the hypothetical mixed state $\rho_{mix}$ given in equation (\ref{eqn:Eqn25}) gives the same cross section as the entangled state $\ket{\Psi_{l}^{+}}$, and therefore would imply identical $R$ values. Furthermore, all ensembles hypothesised by Bohm-Aharonov which possesses the rotational and reflective symmetry of the form given by equation (\ref{eqn:Eqn25}) would evidently have differential cross sections which are distinguishable to the cross section derived for the Bell state $\ket{\Phi^{-}_{c}}$/$\ket{\Psi^{+}_{l}}$.   
\subsectionn{Past experiment}
{ \fontfamily{times}\selectfont
\noindent
The most precise measurements to date of the correlation of two photons corresponding exactly to parapositronium decay are those of Langhoff~\cite{Langhoff1960} and of Kasday et al.~\cite{Kasday1975} who, after geometrical corrections were applied, measured $R=2.47\pm 0.07$ and $R=2.33\pm 0.10$, respectively. Although these results were sufficient to rule out certain hypothetical modifications of quantum mechanics, motivated by Einstein's ideas they did not confirm the predicted ratio of $2.85$ for the annihilation photons.
			
The objective of Kasday et al. was to assume Compton theory of annihilation photons is correct and use them to test quantum theory, local hidden variable theories and the Bohm-Aharonov hypothesis by Compton scattering~\cite{Kasday1971}. The data of Kasday et al. was analysed by computing for each value of the relative azimuthal angle $\eta$ the quantity $R_{exp}(\eta)$ of the form 
\begin{equation}
	R_{exp}(\eta)=\frac{N/N_{ss}}{(n_{1}\{\theta_{1}\}/N_{ss})(n_{2}\{\theta_{2},\eta\}/N_{ss})},
	\label{eqn:Eqn30}
\end{equation}
where
\begin{equation}
	\begin{aligned}
	N_{ss}=&~n_{s1}\cap n_{s2}~\mbox{is the number of times the two photons Compton scatter.}\\
	N=&~n_{1}\{\theta_{1}\}\cap n_{2}\{\theta_{2},\eta\}~\mbox{is the number of times the two photons Compton scatter and both}\\
	&\mbox{photons are detected.}\\
	n_{1}\{\theta_{1}\}=&~\mbox{is the number of times the two photons Compton scatter {\it{and}} only photon 1 is}\\ &\mbox{detected.}\\
	n_{2}\{\theta_{2},\eta\}=&~\mbox{is the number of times the two photons Compton scatter {\it{and}} only photon 2 is}\\
	&\mbox{detected.}\\
	\end{aligned}
	\label{eqn:Eqn31}
	\end{equation} 
and where the quantity $R_{exp}(\eta)$ is taken to be a measure of correlation between the momenta of the scattered photons, such that for uncorrelated photon momenta $R_{exp}(\eta)=1$, and deviations away from 1 correspond to correlations between the momenta.
	
A barrier which still exists for experimentalists is that ideal linear polarization analysers are unavailable for annihilation photons, because of their high energy (511 keV). Nevertheless, using quantum theory, Kasday, Ullman, and Wu~\cite{Kasday1975} showed that the ratio $R_{exp}(\eta)$ is related to an equation for a hypothetical experiment which has access to ideal polarization analysers and which has the functional form
\begin{equation}
	R_{exp}(\eta)=A-B\cos2\eta,
	\label{eqn:Eqn32}
\end{equation}
where $A$ and $B$ are parameters to be fitted to the data. When the measurement of $A=1$ is within the margin of uncertainty, the parameter $B$ can be used to test the validity of  quantum theory. 
			
Equation (\ref{eqn:Eqn32}) can be equated to the theoretical differential cross section of equation (\ref{eqn:Eqn15a}) in the following manner
\begin{subequations}
\begin{equation}
A=1,\quad\mbox{and}\quad B=M_{exp}(\theta_{1})M_{exp}(\theta_{2})\equiv m(\theta_{1})m(\theta_{2}),
\label{eqn:Eqn33a}
\end{equation}
\begin{equation}
\frac{N}{N_{ss}}\equiv\frac{1}{\sigma_{(\gamma_{1}, \gamma_{2})}}\frac{\partial^{2}\sigma^{(\Psi_{l}^{+})}}{\partial\Omega_{1}\partial\Omega_{2}},
\label{eqn:Eqn33b}
\end{equation}
\\
\begin{equation}
\frac{n_{1}\{\theta_{1}\}}{N_{ss}}\equiv\frac{1}{\sigma}\bra{I}T(\theta_{1};1)\ket{I},\quad\mbox{and}\quad\frac{n_{2}\{\theta_{2},\eta\}}{N_{ss}}\equiv\frac{1}{\sigma}\bra{I}T(\theta_{2};1)\ket{I},
\label{eqn:Eqn33c}
\end{equation}	
\end{subequations}
where $M_{exp}$ are experimental factors described in~\cite{Kasday1975}. In terms of an ideal experiment in which the data is obtained using the method of Kasday et al. to compute $A$, $B$ and the ratio $N/N_{ss}$,  these quantities are in principle equivalent to their respective theoretical counterparts, given in equations (\ref{eqn:Eqn33a}) and (\ref{eqn:Eqn33b}). 

However, equations in (\ref{eqn:Eqn33c}) are equated to the terms which are associated with mutually independent events. But no such mutually independent photons exist in an ideal experiment in which pairs of photons are created from the annihilation of parapositronium. Strictly speaking, the theoretical prediction of correlation between the annihilation momenta is just $N/N_{ss}$. Recall that the objective of Kasday was not to measure the differential cross section of annihilation photons, but to assume that the theory of Compton scattering of annihilation photons was correct and to measure $R$ in order to test the aforementioned hypothesis.  
\sectionn{Conclusion}
{ \fontfamily{times}\selectfont
	\noindent
We have shown that the cross section for annihilation photons can be described as the difference between the cross sections of a hypothetical separable state and a hypothetical forbidden maximally entangled state, and that the lobe-like structures correspond to regions where the value of the cross section of the hypothetical forbidden state is maximized.

We derived the cross section for the hypothetical mixed state which was first considered by Bohm and Aharonov and more recently by Hiesmayr and Moskal, and found that the cross section for this state is not identical to the cross section of annihilation photons.We also found a theoretical value of unity for the anisotropy of the mixed state which is consistent with the value calculated by Bohm and Aharonov. We therefore concluded that it is not necessary to reinterpret past experiments that tested the Bohm-Aharonov hypothesis. Also, given that Bohm and Aharonov had already shown that all cross sections for ensembles which possess rotational and reflective symmetry of the form given by the hypothetical mixed state are different to the cross
section for the annihilation photons, we conclude that the lobe-like structures observed in the differential cross section are due to entanglement between the annihilation photons.

To date, no Compton scattering experiment of annihilation photons is known that has directly confirmed the predicted value of $R=2.85$, nor has there been any direct attempt to measure the cross section itself. As a result, the question of the kinematic outcomes of Compton scattering of annihilation photons remains open. New experimental tests are essential to clarify this point. In addition to confirming the predictions of quantum physics, they will also clarify the predictions for the optimal capabilities of Compton PET systems.}

{\color{myaqua}
\vskip 6mm
 \noindent\Large\bf Acknowledgements}
 \vskip 3mm
{ \fontfamily{times}\selectfont
 \noindent
Wewould like to thank Misao Sasaki, Dmitry Khangulyan and Hirokazu Odaka for their helpful discussions. Peter Caradonna gratefully acknowledges the support by JSPS KAKENHI Grant Number JP19K23436, and the continued support of the Centre for Advanced Imaging, the University of Queensland and of the Kavli Institute for the Physics and Mathematics of the Universe (WPI). Tadayuki Takahashi, Shin’ichiro Takeda and Peter Caradonna acknowledge the support by the Grant-in-Aid for Scientific Research on Innovative Areas ‘Toward new frontiers: Encounter and synergy of state-of-the-art astronomical detectors and exotic quantum beams’ and from JSPS KAKENHI grant numbers 16H02170, 18H05457 and 18H02700.}

\bibliographystyle{ieeetr}
\bibliography{BibFilePaper1} 

\end{document}